\documentclass[journal=nalefd,manuscript=letter]{achemso}

\usepackage[version=3]{mhchem} 
\usepackage{gensymb}
\usepackage{xcolor}
\usepackage{textcomp}
\usepackage{amsmath}

\usepackage{caption}

\usepackage[section]{placeins}
\usepackage{graphicx}
\usepackage{amssymb}
\usepackage{amsfonts}
\usepackage{color}
\usepackage[normalem]{ulem}
\usepackage{soul}
\usepackage{epstopdf}
\usepackage{hyperref}
\usepackage{fancyhdr}
\usepackage{bbold}

\author{Simon Baber}
\affiliation[Exeter University]
{College of Engineering, Mathematics and Physical Sciences, University of Exeter, Exeter, EX4 4QF, United Kingdom.}
\author{Ralph Nicholas Edward Malein}
\affiliation[Exeter University]
{College of Engineering, Mathematics and Physical Sciences, University of Exeter, Exeter, EX4 4QF, United Kingdom.}
\author{Prince Khatri}
\affiliation[Exeter University]
{College of Engineering, Mathematics and Physical Sciences, University of Exeter, Exeter, EX4 4QF, United Kingdom.}
\author{Paul S. Keatley}
\affiliation[Exeter University]
{College of Engineering, Mathematics and Physical Sciences, University of Exeter, Exeter, EX4 4QF, United Kingdom.}
\author{Shi Guo}
\affiliation[Exeter University]
{College of Engineering, Mathematics and Physical Sciences, University of Exeter, Exeter, EX4 4QF, United Kingdom.}
\author{Freddie Withers}
\affiliation[Exeter University]
{College of Engineering, Mathematics and Physical Sciences, University of Exeter, Exeter, EX4 4QF, United Kingdom.}
\author{Andrew J. Ramsay}
\affiliation[Hitachi Cambridge]
{Hitachi Cambridge Laboratory, Hitachi Europe Limited, Cambridge, CB3 0HE, United Kingdom}
\author{Isaac J. Luxmoore}
\affiliation[Exeter University]
{College of Engineering, Mathematics and Physical Sciences, University of Exeter, Exeter, EX4 4QF, United Kingdom.}
\email{i.j.luxmoore@exeter.ac.uk}

\title
  {Excited State Spectroscopy of Boron Vacancy Defects in Hexagonal Boron Nitride using Time-Resolved Optically Detected Magnetic Resonance}

\keywords{Color center, hexagonal boron-nitride, 2D materials, optically detected magnetic resonance}

\begin{document}



\begin{abstract}
We report optically detected magnetic resonance (ODMR) measurements of an ensemble of spin-1 negatively charged boron vacancies in hexagonal boron nitride. The photoluminescence decay rates are spin-dependent, with inter-system crossing rates of $1.02~\mathrm{ns^{-1}}$ and $2.03~\mathrm{ns^{-1}}$ for the $m_s=0$ and $m_s=\pm 1$ states, respectively. Time-gating the photoluminescence enhances the ODMR contrast by discriminating between different decay rates. This is particularly effective for detecting the spin of the optically excited state, where a zero-field splitting of $\vert D_{ES}\vert=2.09~\mathrm{GHz}$ is measured. The magnetic field dependence of the time-gated photoluminescence exhibits dips corresponding to the Ground (GSLAC) and excited-state (ESLAC) anti-crossings. Additional dips corresponding to anti-crossings with nearby spin-1/2 parasitic impurities are also observed. The ESLAC dip is sensitive to the angle of the external magnetic field. Comparison to a model suggests that the anti-crossings are mediated by the interaction with nuclear spins, and allow an estimate of the ratio of the spin-dependent relaxation rates from the singlet back into the triplet ground state of $\kappa_0/\kappa_1=0.34$. This work provides important spectroscopic signatures of the boron vacancy, and information on the spin pumping and read-out dynamics.
\end{abstract}

\vspace{5mm}

Hexagonal boron nitride (hBN) is a van der Waals crystal with a wide band gap, and is often used as an insulator in layered two-dimensional devices. Recently, color centers in hBN have attracted considerable attention as room temperature quantum emitters\cite{Sajid_IOPRepProgPhysics2020_BNreview}. Not only can high brightness anti-bunching be observed at room temperature\cite{Tran_NatNano2015,Jungwurth_NL2016_BNtemp}, but for some defect species, the the zero phonon line fraction can be as high as 80\%\cite{Tran_ACSNano2016_BNRobustSPS,Sajid_IOPRepProgPhysics2020_BNreview}, and can exhibit transform-limited linewidths under resonant excitation at room temperature\cite{Hoese_SciAdv2020_FTlimited,Dietrich_PhysRevB2020_BN_FTlimited}. These results suggest promising optical coherence properties that surpass defects found in other wide band gap materials such as diamond\cite{Awschalom_NatPhotonics_review} or silicon carbide\cite{Castelletto_IOP-JPhys_Photonics_review}. More recently, optically detected magnetic resonance (ODMR) experiments have been reported\cite{Gottscholl2020,Gottscholl_SciAdv_VB-CoherentControl,Gottscholl_NatComms_VB-sensing,Liu_arXiv_VB-Rabi,Gao_NanoLett_VB-plasmonic,Kianinia_ACSPhotonics_IonImplantation,Gao_ACSPhotonics_LaserWriting,Liu_ACSPhotonics2021_VBtemperature,Chejanovsky_NatMaterials_BNsingleDefect,stern_arXiv_BNsingle}. These are useful both for identifying the defect species, and for assessing the prospects of emitters for applications such as sensing or spin-photon interfaces.     

So far, ODMR experiments using hBN have mainly focused on ensembles of negatively charged boron vacancies  ($V_B^-$)\cite{Gottscholl2020,Gottscholl_SciAdv_VB-CoherentControl,Gottscholl_NatComms_VB-sensing,Liu_arXiv_VB-Rabi,Gao_NanoLett_VB-plasmonic,Kianinia_ACSPhotonics_IonImplantation,Gao_ACSPhotonics_LaserWriting}, although there have also been reports of experiments with single bright defects of unconfirmed species \cite{Chejanovsky_NatMaterials_BNsingleDefect,stern_arXiv_BNsingle}. Work to date has determined that $V_B^-$ is a radiative spin triplet system (total spin quantum number, $S=1$), and a zero-field splitting of $D_{GS}=+3.45~\mathrm{GHz}$.\cite{Gottscholl2020} So far, the focus has been on the spin properties of the optical ground-state, and nothing is known experimentally about the spin-properties of the excited state, which plays an important role in the spin read-out and pumping process. Furthermore, knowledge of the spin-splittings of the excited state may provide further confirmation of the identity of the defect currently assigned to $V_B^-$.

\begin{figure}
  \includegraphics[width=\textwidth]{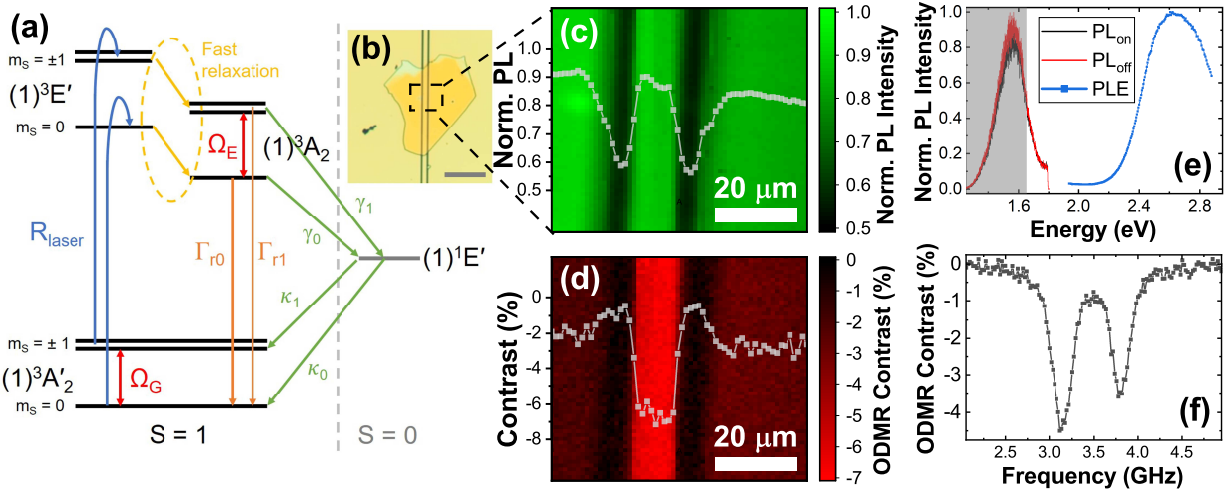}
  \caption{(a) Energy level diagram of the $V_B^-$ defect in hBN (adapted from ref. \citenum{Reimers_PhysRevB_VB-}). (b) Optical micrograph of the coplanar waveguide device with hBN flake on top. The scale marker is 100 $\mathrm{ \mu m}$ (c) Spatial map of the photoluminescence (spectral range from 1.2 to 1.65 eV) from the region indicated by the dashed square in (b). (d) Spatial map of the ODMR contrast for the ground state $m_s=0 \leftrightarrow -1$ transition at a magnetic field of 12 mT, from the same region as in (c). Intensity and contrast profiles along the x-direction, taken from the centre of the images, are overlaid in (c) and (d), respectively. (e) PL and PLE spectra of the $V_B^-$. The gray shaded area indicates the spectral band over which the PL signal is integrated. PL spectra are plotted with and without the microwave excitation. (f) ODMR spectrum of the $V_B^-$ ensemble with a magnetic field of 12 mT.}
  \label{fgr1}
\end{figure}

In this work, we present ODMR experiments with an ensemble of negatively charged boron vacancies, in hBN flakes transferred directly onto microfabricated co-planar waveguides (CPW). The close proximity of the hBN flake to the CPW results in a large ac-magnetic field at the defects. In combination with a time gated detection method, which enhances the ODMR contrast to $\sim30\textnormal{\%}$, we are able to observe resonances corresponding to the spin-levels of the optical excited state. In Fig. \ref{fgr1}(a) an energy level diagram based on the theoretical work of Reimers \textit{et al.}\cite{Reimers_PhysRevB_VB-} is presented. In this picture, the optical cycle in the triplet subspace consists of absorption from the ground state, $(1)^3 A'_2$, which is dominated by the transition to $(1)^3E'$, followed by fast (sub-ps) relaxation to $(1)^3 A_2$ and radiative recombination back to $(1)^3 A'_2$. Using photoluminescence excitation spectroscopy (PLE), we measure this absorption resonance to be at 2.6 eV, consistent with $(1)^3 A'_2\rightarrow(1)^3E'$ transition, as calculated in ref. \cite{Reimers_PhysRevB_VB-}. We measure a zero-field splitting (ZFS) of $\vert D_{ES}\vert=2.1$ GHz for the excited-state spin levels and a g-factor that is similar to the ground state value of $\sim$2. The PL decay rates are found to be spin-dependent, and measured to be $\gamma_0=1.01~\mathrm{ns^{-1}}$, and $\gamma_1=2.03~\mathrm{ns^{-1}}$. This is fast compared with the estimated radiative rate and consistent with estimates of inter-system-crossing times calculated in ref. \cite{Reimers_PhysRevB_VB-}. This spin-dependent relaxation gives rise to the spin read-out contrast. To cross-check the ODMR measurements, the magnetic field dependence of the PL is measured. Dips occur at the expected anti-crossings of the ground and excited states in a microwave-free analog of ODMR. In addition, there are dips that indicate the presence of nearby spin-1/2 defects, that may provide a source of decoherence. Comparison to a model, suggests that the anti-crossings are coupled by the nuclear spins of nearby nitrogen atoms, and the ratio $\kappa_0/\kappa_1=0.34$ is estimated, which determines the spin polarization achieved.

Fig. \ref{fgr1}(b) shows an optical micrograph of our device, which consists of a gold coplanar waveguide (CPW) fabricated on a sapphire substrate. HBN flakes are obtained by mechanical exfoliation from bulk crystal and positioned on the CPW using standard dry-transfer techniques. Boron vacancy defects are introduced/activated using Carbon-ion implantation at 10 keV and a dose of $1 \times 10^{14} \textnormal{ cm}^{-2}$ \cite{Kianinia_ACSPhotonics_IonImplantation}. Photoluminescence (PL) is excited using a broadband supercontinuum laser that is filtered by an acousto-optic tunable filter (AOTF) to give a \(\sim\)1 nm bandwidth and a pulsewidth of \(\sim\)5 ps. All experiments are performed at room temperature and in air. (Further measurement details can be found in the Supporting Information [S.I.]).

The PL spectrum from the implanted flake is shown in Fig. \ref{fgr1}(e), with a broad emission peak centred at around 1.56 eV that is consistent with previous measurements of ensembles of negatively charged boron vacancies\cite{Gottscholl2020}. The emission is uniform across the flake, as shown by the PL intensity map in Fig. \ref{fgr1}(c). The PL intensity is $\sim1.7$ times brighter where the flake overlaps the CPW, compared to the region on the bare sapphire, with the gold acting as a back-reflector to enhance the collection efficiency. The excitation efficiency depends on the energy of the laser, with a PLE spectrum presented in Fig. \ref{fgr1}(e). The absorption peaks at around 2.6 eV, close to the energy of the $(1)^3A_2'\rightarrow (1)^3E'$ transition, predicted by Reimers \textit{et al.}\cite{Reimers_PhysRevB_VB-} to dominate the absorption. In the following experiments the laser is tuned to 2.6 eV to match the peak of this absorption.

In the case of continuous wave (CW) ODMR, a microwave signal is modulated on and off, and the contrast (defined as $(I_{on} - I_{off})/I_{off}$ where $I_{on}$ and $I_{off}$ are the PL intensity with the microwaves on and off, respectively) is measured. A typical CW-ODMR spectrum is plotted in Fig. \ref{fgr1}(f) for a magnetic field of 12 mT. The two resonances at $\sim3.1$ GHz and $\sim3.8$ GHz correspond to the $m_S = 0$ to $m_S = -1$ and $m_S = 0$ to $m_S = +1$ transitions in the optical ground state, as previously reported\cite{Gottscholl2020}. The ODMR contrast of the $m_S = 0$ to $m_S = -1$ transition is mapped across the sample in Fig. \ref{fgr1}(d), with strong contrast only observed from the central conductor of the CPW, where the ac-magnetic field has a strong in-plane component. 

\begin{figure}
  \includegraphics[width=\textwidth]{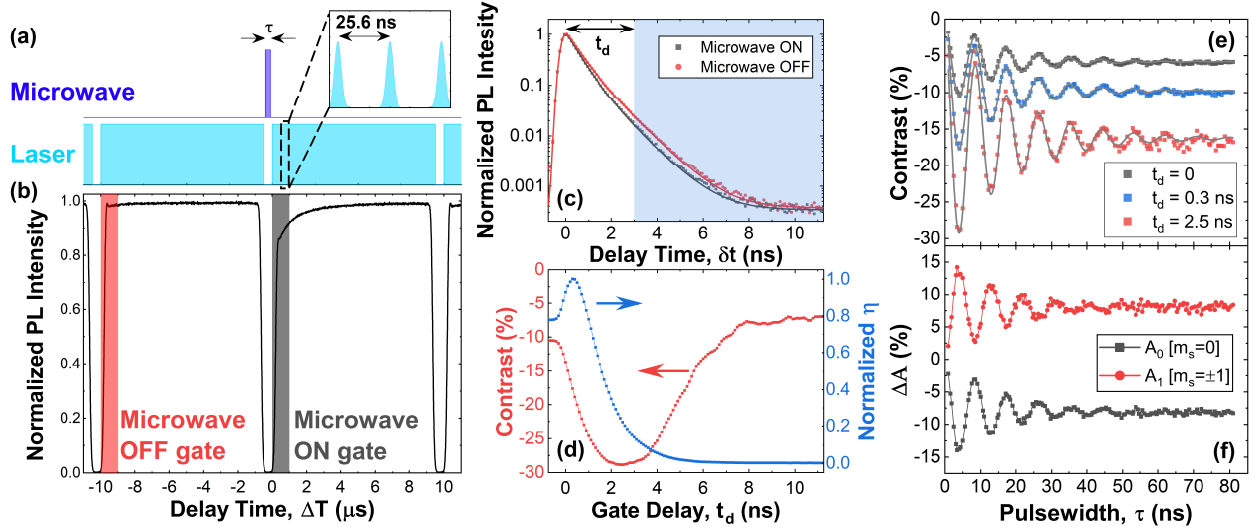}
  \caption{Enhanced contrast of ODMR using time filter. (a) Pulse sequence for time-resolved ODMR measurements. (b) Typical time-averaged PL intensity trace resulting from pulse sequence in (a), for time-binning of photons by their arrival time relative to the microwave pulse, $\Delta T$. The red and gray shaded regions indicate the collection windows used to plot time-resolved PL traces. (c) Time-resolved PL traces for time-binning of photons by their arrival time relative to the laser pulses, $\delta t$. The black and red cures are for photons detected during the microwave on and off windows shown in (b), respectively. The gate delay time, $t_d$, is defined from the peak of the time-resolved PL trace, with the blue shaded region indicating the photons which are used to calculate the contrast plotted in (d). (d) Time-resolved contrast and sensitivity figure of merit, $\eta$ for a microwave pi-pulse. (e) ODMR contrast as a function of microwave pulsewidth, $\tau$, showing Rabi-oscillation of the $m_S = 0$ to $m_S = -1$ ground state transition. At each pulsewidth time-resolved traces are recorded and the contrast is calculated for the three different gate delay times indicated. (f) From the same data set as (e) exponential amplitudes are extracted from bi-exponential fits to the time-resolved PL curves and the change in these coefficients between when the microwave pulse is on and off, $\Delta A_{0,1} = \left(A_{0,1(on)}-A_{0,1(off)}\right)/\left(A_{0(off)}+A_{1(off)}\right)$, is plotted. In all panels the external magnetic field is 12 mT.}
  \label{fgr2}
\end{figure}

In previous work with NV centers in diamond\cite{Batalov_PRL_2008_NVTRPL} and divacancies in SiC \cite{Klimov_SciAdv_SiC}, the spin-dependent lifetimes of the time-resolved photoluminescence was used as a contrasting agent. We now trial this method for the $V_B^-$ in hBN. In Fig. ~\ref{fgr2}(a) the basic pulse-sequence is illustrated. To optically excite the sample, the 39 MHz pulse train from the spectrally filtered supercontinuum is switched on and off with an acousto-optic modulator. In this case, with a repetition frequency of 100 kHz and a duty-cycle of 95 \%. The $9.5 \mu $s long train of pulses optically pump the ensemble of boron vacancies into the ground $m_S = 0$ state \cite{Gottscholl2020}. To rotate the spin of the ground-state, synchronized microwave pulses are applied at 50 kHz, half the optical repetition frequency. Fig. \ref{fgr2}(b) shows the photoluminescence signal when the photons are time-binned according to their arrival time relative to the 50 kHz repetition frequency of the microwave pulses, $\Delta T$. This is averaged over many repetition periods of the laser, which is not time synchronised with the 50 kHz frequency of the microwave pulses. This results in the PL trace shown in Fig.  \ref{fgr2}(b), which is equivalent to that recorded with a CW laser. A clear reduction in the intensity is observed following a microwave pi-pulse. The signal then recovers, as the laser optically pumps the system back into the $m_s=0$ state.

From this data set it is also possible to analyse the decay of the optical transitions. Time-resolved PL (TRPL) traces are made by time-binning the photon arrival times, $\delta t$ relative to the 39 MHz repetition rate of the laser.  The PL is averaged over the $\sim 39$ pulses that lie within the $1~\mathrm{\mu s}$ long windows highlighted in Fig. \ref{fgr2}(b). Such TRPL traces are plotted in Fig. \ref{fgr2}(c) with and without a microwave pi-pulse. The PL counts for both traces are normalized to the maximum value of the microwave OFF trace. Applying the microwave pulse results in an overall reduction in intensity, and a speed up in the PL decay. The TRPL curves of Fig. \ref{fgr2}(c) are fit with a bi-exponential decay of the form $A_0 e^{-t\gamma_0} + A_1 e^{-t\gamma_{1}}$ (convolved with a Gaussian of 330 ps FWHM to account for the detector response), PL decay  rates are found to be $\gamma_0 = 1.01 \hbox{ ns}^{-1}$ and $\gamma_{1} = 2.03 \hbox{ ns}^{-1}$. Applying a microwave pi-pulse, enhances the fast ($A_1$), and suppresses the slow ($A_0$) component of the PL decay. If optical pumping initializes the ground state into $m_s=0$ state \cite{Gottscholl2020}, this implies that the fast decay can be assigned to $m_s=\pm 1$, and the slow decay to $m_s=0$ of the excited state. In ref. ~\citenum{Reimers_PhysRevB_VB-}, the relaxation of the excited state of the optical transition, $(1)^3A_2$, was predicted to be dominated by an inter-system crossing to $(1)^1E'$. The ISC-lifetime was calculated to be 1.7 ns at room temperature, compared to a radiative lifetime of 11~$\mathrm{\mu s}$, in reasonable agreement with our measurements. This spin-dependence of the inter-system crossing rates gives rise to the spin-read out of the ODMR.

Since the PL decay rates are spin dependent, the ODMR contrast can be enhanced by selecting only the photons that arrive after a gate-time, $t_d$, with respect to the laser pulse. In Fig. \ref{fgr2}(d), the contrast, $C(t)=\left[I_{ON}(t)-I_{OFF}(t)\right]/I_{OFF}(t)$ is plotted vs the time-delay $t_d$, where $I_{ON}$ and $I_{OFF}$ are the integrated counts from $t_d$ to $t_{rep}/2$ for the microwave ON and OFF traces, respectively, and $t_{rep}$ is the repetition period of the laser. At $t_d=0$, the full signal is collected and the contrast for a microwave  $\pi$-pulse is $\sim10$\%. The contrast increases with $t_d$, reaching an optimum of almost 30\% at $t_d\approx2.5~\mathrm{ns}$, a $\sim$ 3-fold enhancement. To quantify the improvement in sensitivity of the time-gated readout,  we calculate a figure of merit, $\eta(t)=\left(I_{ON}(t)+I_{OFF}(t)\right)C(t)^2$, which is plotted as a function of $t_d$ in Fig. \ref{fgr2}(d). This is proportional to the sensitivity of an ODMR B-field sensor \cite{Barry_RevModPhys_NVMagnetometry}, and shows a modest improvement of $\sim20$\% at $t_d=0.3~\mathrm{ns}$. To illustrate the enhancement in contrast due to time-gating, Fig. \ref{fgr2}(e) compares Rabi oscillation measurements with a gate delay of $t_d=0$ (no time-gating), $t_d=0.3~\mathrm{ns}$ (maximum $\eta$) and $t_d=2.9~\mathrm{ns}$ (maximum contrast).  A Rabi oscillation measurement can also be made by plotting the relative change in the amplitudes of the exponential components $A_0$, and $A_1$ vs microwave pulse width, $\tau$ (Fig. \ref{fgr2}(f)). Note that the change in amplitude is equal and opposite ($\Delta A_0\approx -\Delta A_1$), indicating the transfer of population from $m_s=0 \leftrightarrow -1$, and confirming that the read-out contrast is arising from the spin-dependent decay rates.

\begin{figure}
  \includegraphics[width=\textwidth]{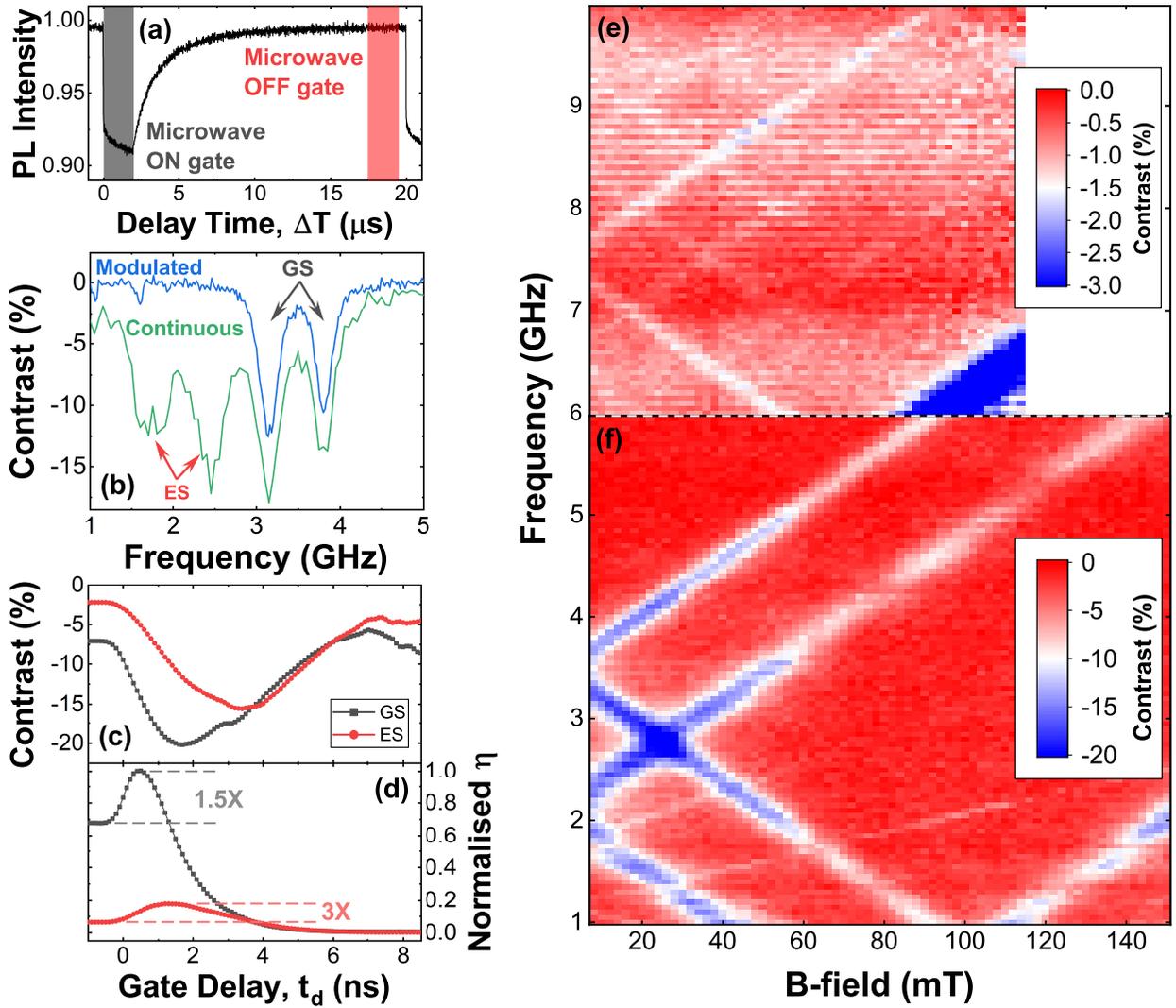}
  \caption{Zeeman splitting of ground and excited states of the negatively charged boron vacancy. (a) Time averaged PL intensity resulting from a $2 \mu s$ microwave pulse with a repetition rate of 50 kHz. (b) Comparison of ODMR spectra for microwave excitation applied with and without the laser applied. (c) and (d) Time resolved contrast (c) and sensitivity figure of merit, $\eta$ (d) for microwave excitation of the ground ($m_S = 0$ to $m_S =-1$ [3.14 GHz]) and excited ($m_S = 0$ to $m_S =+1$ [2.44 GHz]) states. The curves in (d) are normalized to the maximum of $\eta$ for the ground state transition. (e) and (f) Color plots of the ODMR contrast as a function of microwave frequency and external magnetic field for $t_d=3.4 ~\mathrm{ns}$. In (e) the contrast scale is limited to -3 \% to highlight the transition with $\sim7.5$ GHz ZFS. In (a) - (d) the external magnetic field is 12 mT.}
  \label{fgr3}
\end{figure}

The close proximity of the hBN flake to the CPW enables a strong ac-magnetic field, and a large Rabi-frequency reaching $\sim110 \hbox{ MHz}$, limited by the maximum available input power of +30 dBm (see S.I.). The combination of this strong ac-field and the time-resolved detection method allows us to detect the spin structure of the excited state, despite the very fast ISC times. Figure \ref{fgr3}(a) presents a PL time-trace, with the PL averaged over a time window of $\sim$39 pulses, with the pump laser always on. Following resonant microwave excitation of the ground state $m_s=0\leftrightarrow -1$ transition, there is a rapid drop in PL signal, which recovers when the microwaves are switched off. This indicates that the microwave excitation is strong enough to overcome the optical pumping. Figure \ref{fgr3}(b) compares ODMR spectra taken with the laser on/off during microwave excitation, using the time-gated method, at a B-field of 12 mT. When the laser is off, only two Zeeman-split peaks centered on the ZFS $D_{GS}\approx 3.5$~GHz, matching the grounds-state are observed \cite{Gottscholl2020}. However, if the laser is on during microwave excitation, a second excited state doublet with ZFS $D_{ES}\approx 2.1~\mathrm{GHz}$ is also observed. Fig. \ref{fgr3}(e,f) present a color-map of the ODMR spectra vs B-field. A third, much weaker, transition with ZFS of $\sim7.5$ GHz is also observed.

Interestingly, the contrast vs gate-time reveals different behavior for the excited and ground state transitions. In Fig. \ref{fgr3}(b), the contrast is plotted versus gate-time for excitation frequencies corresponding to the ground ($m_S = 0$ to $m_S =-1$ [3.14 GHz]) and excited ($m_S = 0$ to $m_S =+1$ [2.44 GHz]) state transitions with a magnetic field of 12 mT. With no time gating the ground state contrast is -7.1 \% and decreases rapidly with gate delay time, reaching a maximum negative contrast of $-20 \%$ at $t_d=1.7 ~\mathrm{ns}$. An increase of 2.8 times, similar to the case of pulsed ODMR, in the laser off case, shown in Fig. \ref{fgr2}. For the excited state, the contrast with no time gating is $-2.1 \%$ and the initial rate of decrease is slower than for the ground state, but the rate then increases and reaches a maximum contrast of $-16 \%$ at $t_d=3.4 ~\mathrm{ns}$, an enhancement of 7.5 times. As shown in Fig. \ref{fgr3}(c) this translates to different gate delay times for optimum read-out sensitivity of the ground and excited states. For the ground (excited) state the read-out sensitivity is increased by 1.5 times at $t_d=0.5 ~\mathrm{ns}$ (3 times at $t_d=3.4 ~\mathrm{ns}$). The time gated detection therefore plays a crucial role in resolving the excited state transitions, as highlighted in Fig. \ref{fgr_SI2} of the S.I and hence the data in Fig. \ref{fgr3}(f) is plotted for $t_d=3.4 ~\mathrm{ns}$. 

The difference in the time evolution of the contrast for the ground and excited states can be understood with reference to the energy level diagram in Fig. \ref{fgr1}(a). In the case of the ground state resonance, contrast arises from a difference in the ground-state $m_s=0$ and $m_s=\pm1$ populations, which determines the weighting of the two exponential amplitudes in the PL decay. These weights are determined by microwave excitation of the ground state, therefore there is no change in these weights during the decay of the PL. In the case of the excited state, the laser pulse transfers the initialised spin populations to the excited state, at which point the microwave excitation at $\Omega_E$ begins a Rabi-oscillation. This results in transfer of population between $m_s=0$ and $m_s=\pm1$ during the PL decay leading to a time trace with oscillatory character \cite{Fuchs_NatPhys_NVESRabi}. We are unable to resolve the Rabi-oscillation of the excited state in our measurements because the maximum available Rabi-frequency of $\sim105$ MHz is much less than the PL decay rate, $\left( \gamma_0+\gamma_1 \right)/2 = 1.5 ~\mathrm{GHz}$ (see S.I. for modelling with the Quantum Toolkit in Python [QuTip]\cite{JOHANSSON20121760,JOHANSSON20131234}).  

\begin{figure}
  \includegraphics[width=\textwidth]{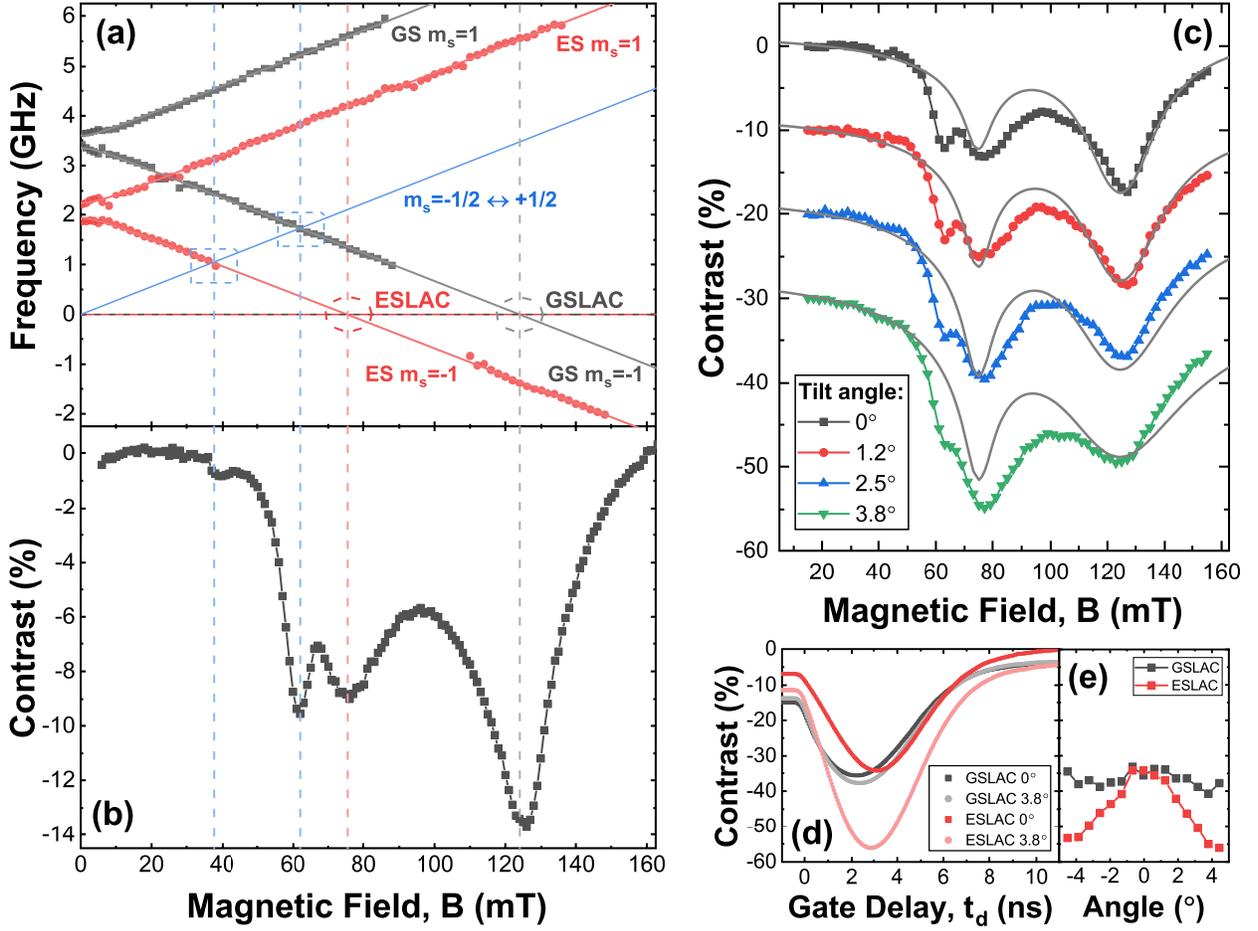}
  \caption{Level anticrossings of the negatively charged boron vacancy. (a) Frequencies of the ground and excited state resonances extracted from Gaussian fits to the data in Fig. \ref{fgr2}(d). The solid red (gray) lines show fits to the data of Eq. \ref{eqn1} for the excited (ground) state transitions. The red (gray) vertical dashed lines indicate the magnetic field corresponding to the ESLAC (GSLAC). The solid blue line indicates the frequency difference of a $S=1/2$ impurity and the dashed blue lines indicate the frequencies at which this anticrosses with the $V_B^-$ ground and excited states. A positive sign of $D_ES$ is assumed. (b) Contrast of the time resolved photoluminescence (defined as $(I_B - I_{ref})/I_{ref}$, where $I_B$ is the PL intensity at the given field and $I_{ref}$ is the reference PL intensity, recorded at 20 mT) as a function of the applied magnetic field. Dips in the intensity at $\sim76$ mT and $\sim125$ mT correspond to the excited state and ground state level anticrosings, respectively. (c) Magnetic field dependence of the contrast for different tilt angles of the magnetic field relative to the c-axis of the hBN. The curves are offset by -10\% for clarity. The solid gray lines show spectra calculated using the model described in the main text and S.I. In (b) and (c) $t_d=0.5~\mathrm{ns}$ (d) Time-resolved contrast at the excited and ground state level anticrossings at two tilt angles of the external magnetic field. (e) Plot of the maximum contrast at the ESLAC and GSLAC versus the tilt angle of the magnetic field.}
  \label{fgr4}
\end{figure}

The resonant frequencies of the ground and excited state transitions are extracted for each magnetic field and plotted in Fig. \ref{fgr4}(a), which also shows fits to the data of:
\begin{equation}
    \nu_{-,+}=\frac{1}{h} \left( D \pm \sqrt{E^2 + \left(g \mu_B B\right)^2} \right) \label{eqn1}
\end{equation}
where D and E are the ZFS parameters, $\mu_B$ is the Bohr-magneton and $g$ the electron g-factor\cite{Gottscholl2020}. From the fits to the data we find values of $D_{GS}/h=3.47$ GHz, $E_{GS}/h=126$ MHz and $g_{GS}=2.001$ for the ground state, in broad agreement with previous reports\cite{Gottscholl2020,Gottscholl_SciAdv_VB-CoherentControl,Gottscholl_NatComms_VB-sensing,Liu_arXiv_VB-Rabi,Gao_NanoLett_VB-plasmonic,Kianinia_ACSPhotonics_IonImplantation,Gao_ACSPhotonics_LaserWriting}, and $\vert D_{ES}/h\vert=2.09$ GHz, $\vert E_{ES}/h\vert=154$ MHz and $g_{ES}=1.980$ for the excited state. The excited state resonances are broad, and independent of the external magnetic field, with a full-width half maximum (FWHM) of $\sim400$ MHz. This is roughly consistent with the FWHM expected from the excited state lifetime, $FWHM = \left(\gamma_0 + \gamma_{\pm 1}\right)/ \pi = 0.97$ GHz \cite{Fuchs_ES_NV_PRL}.

Our experimental results are consistent with the theoretical work of Reimers \textit{et al}\cite{Reimers_PhysRevB_VB-}. We assign the ES with ZFS of 2.09 GHz to the $(1)^3A_2$ level, as this is the excited state of the observed optical transition. The origin of the transition with ZFS of 7.5 GHz is uncertain. It could originate from another, unidentified defect species within our sample, but it seems more plausible that it is related to an energy level within the $V_B^-$. In this case, the most likely candidates would be $(2)^3A_2'$ or $(1)^3A_1''$. However, at room temperature both states are predicted to relax to $(1)^3A_2$ on timescales less than 1 ps. Fast relaxation would account for the weakness of the ODMR contrast, but it seems unlikely that our experiments could resolve a process on this timescale. Further experiments combining temperature dependence, PLE and ODMR could shed further light on the origin of this transition.

Fig. \ref{fgr4}(a) also illustrates that anticrossings of the $m_s=0$ and $m_s=-1$ states are expected at $\sim76$ mT and $\sim125$ mT, for the excited and ground states, respectively. These anticrossings are investigated by measuring the TRPL as a function of the external magnetic field, with no microwave excitation, see Fig. \ref{fgr4}(b). Here, the contrast is defined as $(I_B - I_{ref})/I_{ref}$, where $I_B$ is the PL intensity at the given field and $I_{ref}$ is the reference PL intensity recorded at B=20 mT. Dips in the contrast are observed at 76 mT and 125 mT, as expected for spin mixing at the ESLAC and GSLAC, respectively. In addition, two further minima can be seen in Fig. \ref{fgr4}(b), a strong feature at $\sim62$ mT and a weaker one at $\sim38$ mT, which correspond to half the magnetic field of the GSLAC and ESLAC, respectively. As shown in Fig. \ref{fgr4}(b), the Zeeman splitting of a spin-1/2 system is resonant with the  $m_s= 0 \leftrightarrow -1$ transition in the excited (ground) state of $V_B^-$ at a magnetic field of $\sim38$ ($\sim62$) mT. Therefore, we propose that these features originate from a coupling of the boron vacancies to neighbouring paramagnetic defects of another species or charge state, as seen previously for NV centres coupled to nitrogen\cite{Hanson_PhysRevLett_NV-N-CoupledSpins} and P1\cite{Anishchik_PRB_NVP1_2017} defects in diamond. The spin dephasing time $T_2^*\approx19 ~\mathrm{ns}$ (extracted from Rabi-oscillation data, see S.I.) and $T_1=10~\mathrm{\mu s}$ are shorter than previous reports \cite{Gottscholl_SciAdv_VB-CoherentControl,Liu_arXiv_VB-Rabi,Gao_NanoLett_VB-plasmonic}, which could also indicate high levels of impurities. As our samples are implanted using C ions it is possible these are carbon related. However, PL measurements show no signature of C-related defects \cite{Mendelson_NatMater2021_Carbon} and indicate that boron vacancies are the only luminescent defect present in large concentrations. Experiments using samples fabricated with different ion-species\cite{Kianinia_ACSPhotonics_IonImplantation} and/or laser irradiation\cite{Gao_ACSPhotonics_LaserWriting}, could shed further light on the nature of these defects.

Finally, by tilting the sample with respect to the field axis of the permanent magnet, we investigate the angular dependence of the external magnetic field on spin-level mixing in the excited and ground states. Fig. \ref{fgr4}(c) plots the contrast versus magnetic field for four different tilt angles. As the angle increases so does the contrast at the ESLAC, whilst the contrast at the GSLAC is relatively unaffected (see also Fig. 4(e)). To gain some understanding of these microwave-free ODMR spectra a model is studied based on the energy level diagram shown in Fig. \ref{fgr1}(a). The model considers 7 energy-levels, 2x3 for the S=1 ground ($(1)^3A_2' $), and excited ($(1)^3A_2$) states, and a shelving state ($(1)^1A_1$). The relaxation rates are as depicted in Fig. \ref{fgr1}(a). Following ref. \citenum{Epstein_NatPhys2005_Anisotropic}, the anti-crossing of the states $m_s=0$, $m_s=-1$ is treated as a Bloch-vector that rotates under an external, and an internal B-field. The optical pumping preserves the spin-z component. (A full description of the model can be found in the S.I.). A comparison of the model to data can be found in Fig. \ref{fgr4}(c). Since the radiative rate is slow compared with the inter-system-crossing rates, ($\Gamma_R\approx 1/(11~\mathrm{\mu s})\ll \gamma_0,\gamma_1$) \cite{Reimers_PhysRevB_VB-}, the read-out contrast is largely determined by the measured values of $\gamma_0,\gamma_1$ and the gate delay time. The dip at $\sim125$ mT arises due to the erasure of the spin initialization by the GSLAC. Since, the contrast is a ratio, the GSLAC dip is determined by the ratio $\frac{\kappa_0}{\kappa_1}\approx 0.34$, and is independent of the absolute value of $\kappa$, which we arbitrarily fix as $\kappa_0=20~\mathrm{\mu s^{-1}}$, since in theory it is expected to be slow \cite{Reimers_PhysRevB_VB-}. The width of the GSLAC is determined by the in-plane effective B-field.  An internal in-plane B-field of 140 MHz, that is perpendicular to the external in-plane B-field, gives a good match to data. 

The ESLAC dip at $\sim76$ mT is harder to model. The dip arises from the effect of the ESLAC on the read-out process. At time zero, the excited state spin is aligned along spin-z just like the reference signal. However, near the ESLAC the spin rotates enhancing the contrast in the PL signal. This makes the time-gating especially helpful for detecting the ESLAC (see Fig. \ref{fgr4}(d)). In contrast to the GSLAC dip, the ESLAC dip is determined by the in-plane effective B-field, and the damping of the spin precession. The change in the ESLAC with tilt is large, and can only be explained by an internal in-plane B-field of approximately 140 MHz that is co-aligned with the external in-plane B-field, with damping that is lifetime-limited by the ISC rates $\gamma_0,\gamma_1$. 

The nearest neighbors of a boron vacancy are three nitrogen atoms, with nuclear spin $I(^{14}N)=1$, and a hyperfine constant of $A=47~\mathrm{MHz}$ \cite{Gottscholl_NatComms_VB-sensing}. This suggests a maximum Overhauser field of 141 MHz. In Fig. \ref{fgr4}(c), the model curves are calculated assuming the internal B-field is due to the nuclear spins only. For each value of $-3\leq m_I\leq3$, a spectra is calculated for an internal $B_z=Am_I$, and $\vert B_{xy}\vert=\sqrt{I(I+1)-m_I^2}$. The resulting spectra is calculated as a sum weighted by the degeneracy of the $m_I$ state to yield a spectra of randomly oriented nuclear spins. The overall match to data, where only the ratio $\frac{\kappa_0}{\kappa_1}=0.34$ is adjustable is good, with the disagreement regarding the width of the ESLAC dip possibly connected to a nuclear pumping effect that is responsible for orienting the in-plane direction of the Overhauser field.

In conclusion, an ODMR study of negative boron vacancies in hBN is reported. The spin dependent inter system relaxation rates responsible for the spin read-out are measured to be $\gamma_0=1.01~\mathrm{ns^{-1}}$, and $\gamma_1=2.03~\mathrm{ns^{-1}}$ for $m_s=0$,$m_s=\pm 1$, respectively. Using a time-gated detection scheme to enhance the ODMR contrast, the zero-field splitting of the excited state is measured to be $D_{ES}=2.09~\mathrm{GHz}$. The magnetic resonances can also be observed in magnetic field dependent PL measurements, from which the ratio of the spin-dependent relaxation rates from the singlet sub-space back into the triplet ground state can be estimated. These are key parameters for understanding the spin initialization and read-out behind the ODMR contrast. Questions remain over how the system escapes the singlet sub-space, since Reimers \textit{et al.}\cite{Reimers_PhysRevB_VB-} predict that $\kappa$ could be extremely slow, and it is possible that the reset process requires optical excitation. Characterization of the excited state spin-structure is also important as a prospective route to optically initialize the nuclear spins\cite{Jaques_PRL_NP_2009,Fischer_PRB_NPatESLAC_2013,Smeltzer_PRA_NP,Falk_PRL_SiC,Klimov_SciAdv_SiC}, an important challenge for this system, if it is to be used as a resource in quantum control and sensing experiments and applications\cite{Klimov_SciAdv_SiC}. 

\newpage

\section{Supporting Information}

\section{1. Experimental Setup}

Photoluminescence is excited using a broadband supercontinuum laser that is filtered by an acousto-optic tunable filter (AOTF) to give a \(\sim\)1 nm bandwidth and a pulsewidth of \(\sim\)5 ps. The laser is equipped with a pulse-picker to adjust the repetition frequency, which in this work is set to either 39 or 78 MHz. The laser is coupled to a long working distance objective lens, with numerical aperture of 0.8, which focuses the light to a diffraction-limited spot $< 1\mu$m in diameter. The luminescence is collected with the same objective and coupled to either a monochromator/CCD or, via a 750 nm long pass filter, to a fiber coupled single photon avalanche diode (SPAD) to enable photon counting and time-resolved fluorescence measurements with a time-tagging module. To perform ODMR experiments an in-plane ac-magnetic field is generated using a microwave waveform from an arbitrary waveform generator that is amplified and applied to the CPW. A permanent magnet mounted on a translation stage below the sample provides the out-of-plane magnetic field.

\section{2. Rabi-oscillations}

\begin{figure}
  \includegraphics[width=\textwidth]{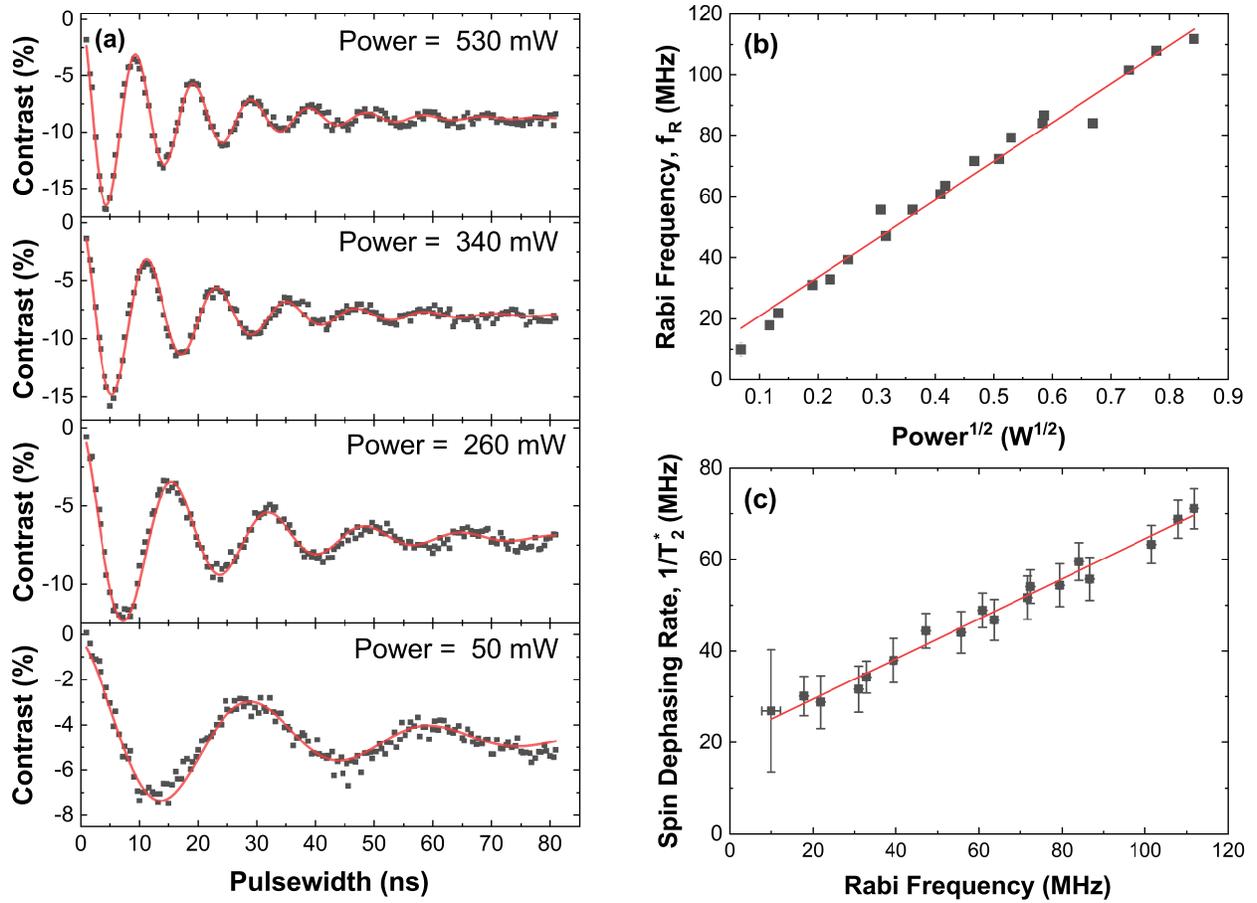}
  \caption{(a) Rabi-oscillation data and fits for different microwave input powers. (b) Dependence of Rabi-frequency on the microwave power. (c) Spin dephasing rate vs Rabi frequency. The x(y)-error bars indicate the standard error of the fit parameters. The red line is a linear fit to the data. In all panels the gate delay time $t_d=0.5 ~\mathrm{ns}$.}
  \label{fgr_SI1}
\end{figure}

Rabi-oscillation data is presented in Fig. 2 of the main paper. This experiment is repeated for different microwave powers and the data fit to $A + B\exp{\left(-\tau/2T_2^*\right)}\cos\left(2 \pi f_R \tau + \theta \right)$. Examples of the Rabi-oscillation data and fits are shown in Fig. \ref{fgr_SI1}(a). The Rabi frequency, $f_R$ extracted from the fits is plotted as a function of microwave power in Fig. \ref{fgr_SI1}(b). At the maximum available power, the Rabi frequency is 112 MHz. The spin dephasing rate, $1/T_2^*$, is plotted in Fig. \ref{fgr_SI1}(c) and shows a linear dependence on the Rabi frequency, indicating power broadening. The dephasing time at the lowest power is $\sim19 ~\mathrm{ns}$, which is 2-3 times shorter than previous reports \cite{Gottscholl_SciAdv_VB-CoherentControl,Gao_NanoLett_VB-plasmonic,Liu_arXiv_VB-Rabi}.

\section{3. Time-gated ODMR Contrast}

\begin{figure}
  \includegraphics[width=\textwidth]{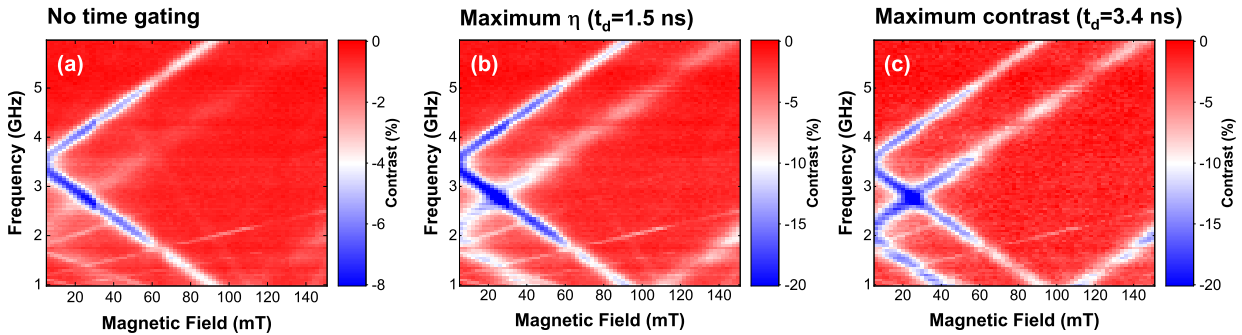}
  \caption{Color plots of the ODMR contrast as a function of microwave frequency and external magnetic field for (a) no time gating and gate delay times (b) $t_d=1.5 ~\mathrm{ns}$ and (c) $t_d=3.4 ~\mathrm{ns}$.}
  \label{fgr_SI2}
\end{figure}

As discussed in the main text, the ODMR contrast is dependent on the gate delay time, $t_d$. In Fig. \ref{fgr_SI2} the dependence of the contrast on magnetic field and frequency are plotted for the case of no time gating and for two different gate times. In all three cases the ground state resonances are well-resolved. However, the excited state resonances are clearly observed only when time gating is applied

\section{4. Excited State contrast modelling and Rabi Frequency}

\begin{figure}
  \includegraphics[width=\textwidth]{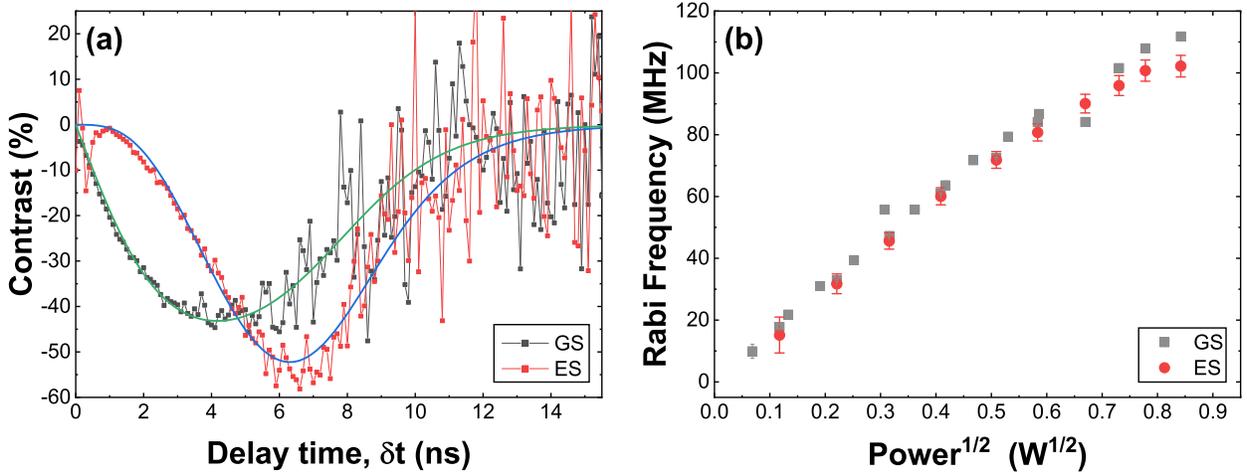}
  \caption{(a) Time resolved contrast plots for the ground and excited state transitions at a magnetic field of 12 mT. The excitation frequency is 2.44 GHz (3.14GHz) for the excited (ground) state measurement. The solid green and blue lines show fits to the data of the model described in the S.I. text. (b) Power dependence of the excited state Rabi-frequency as extracted from the model fits. The ground state Rabi-frequency from Fig. \ref{fgr_SI1}(b) is also plotted for comparison}
  \label{fgr_SI3}
\end{figure}

To better understand the discrepancies between time-resolved contrast plots with microwave applied to excited and ground state levels, the system was modelled using a Lindblad master equation formulation and the model was used to fit the data. This was achieved using the Quantum Toolkit in Python (QuTIP) package\cite{JOHANSSON20121760,JOHANSSON20131234}. The model corresponds to the level diagram in Fig. 1(a) of the manuscript, with the simplification that the fast relaxation between ${}^3E'$ and ${}^3A_2$ is ignored and the excited state is treated as one system of 3 $m_S$ levels. Optical excitation, radiative relaxation and inter-system crossings are represented by incoherent collapse operators. To calculate the contrast, normalised time-resolved PL traces are extracted by finding the time-varying populations of the excited radiative states and adding them to an experimentally-determined background. This is done for both microwave on and microwave off cases, and then the contrast is given as the sum of these two cases, divided by the values with microwave off. This gives an ``instantaneous'' contrast at each delay time in the time resolved plot (see. Fig. \ref{fgr_SI3}(a)).

The data was fit using this model for a range of microwave powers. The fit parameters are the intersystem crossing rates from triplet to singlet, the initial populations of the $m_S$ levels with no microwave applied, and the amplitude and background of the TRPL plots that are used to calculate the contrast. These are kept the same across all microwave powers. Other fit parameters are dependent on whether the microwave is applied to the excited (ES) or ground (GS) state $m_S$ levels. When applied to the GS, it is assumed that the optical pumping transfers the initial population of the $m_S$ levels in the ground state directly into the excited state, so the initial state of the system at time zero is in the excited state. When the microwave is on, the $m_S$ levels are affected before the optical excitation, so the populations after application of microwave and optical pumping are fit parameters that vary with microwave power. By contrast, in the ES case, the microwave is applied during and after the optical excitation pulse. In this case, both the optical pulse and microwave must be modelled. The laser pulse length is given by the experimental conditions, and the optical pump rate is a free parameter, though the same across all microwave powers. The microwave is modelled in the Hamiltonian as a coherent transfer between $m_S = 0$ and $\pm1$ levels, where the Rabi frequency is a free parameter for each microwave power. 

In order to ensure agreement between ISC rates and the state after spin initialisation, the GS and ES cases are fit simultaneously for a range of microwave excitation powers, with example fits shown in Fig. \ref{fgr_SI3}(a). The model shows good agreement with the data, especially with the discrepancy between GS and ES at short times. From the fits the excited state Rabi frequency can be extracted and is found to be similar to the case of the ground state with a maximum of $\sim105$ MHz (Fig. \ref{fgr_SI3}(b)).

\section{5. Modelling of microwave-free ODMR Spectra}

The model is depicted in Fig. 1(a) of the main paper. In total, there are 7 energy-levels. Three for the $S=1$ ground ($(1)^3A_2'$) and excited ($(1)^3A_2$) state, and a single shelving state, $p$ in the singlet space ($^1E'$). To simplify the model, the coherence is only retained for the admixed states $m_s=0, m_s=-1$, which are described by a Bloch-vector $\mathbf{S}$. In addition, the total population on the $m_s=0$ and $m_s=-1$ states is given by $N$, and the remaining population in $m_s=+1$ is given by $N_{+1}$ for a total of 5+5 states describing the ground (lower case) and excited states (upper case) for a total of 11 states such that the vector $\mathbf{x}=(\mathbf{S},N,N_{+1},p,\mathbf{s},n,n_{+1})$ describes the system.

The Bloch-vector is acted on by an effective magnetic field $\mathbf{\Omega}=(\Omega_x,0,\Omega_z)$, where $\Omega_z=D_{es}-g\mu_B B_z$ is shifted by the crystal-field splitting. The other terms are the relaxation rates depicted in fig. 1(a). The resulting first-order rate-equations can be summarized by the matrix $\mathbf{M}$ as $\mathbf{\dot{x}}=\mathbf{M}\mathbf{x}$.

In full the equations are:
\begin{eqnarray*}
\dot{S}_x=-\Gamma_{2es}S_x-\Omega_xS_z \nonumber\\
\dot{S}_y=\Omega_zS_x-\Gamma_{2es}S_y-\Omega_xS_z \nonumber\\
\dot{S}_z=\Omega_xS_y-(\Gamma_R+\Gamma_{1es}+\bar{\gamma})S_z -\Delta\gamma N \nonumber \\
\dot{N}=-\Delta\gamma S_z-(\Gamma_R+\bar{\gamma})N \nonumber \\
\dot{N}_{+1}=-(\Gamma_R+\gamma_1)N \nonumber \\
\dot{p}=\Delta\gamma S_z+\bar{\gamma}N+\gamma_1N_{+1}-(\kappa_0+2\kappa_1)p \nonumber \\ 
\dot{s}_x=-\Gamma_{2gs}s_x-\omega_zs_y \nonumber \\
\dot{s}_y=\omega_zs_x-\Gamma_{2gs}s_y-\omega_xs_z \nonumber \\
\dot{s}_z=+\Gamma_R(S_z/2-N/2+N_{+1})+\Delta\kappa p\nonumber \\ +\omega_xs_y-\Gamma_{1gs}(s_z-n-n_{+1}) \nonumber \\
\dot{n}=\Gamma_R(-S_z/2+N/2+N_{+1})+(\kappa_0+\kappa_1)p+\Gamma_{1gs}n_{+1} \nonumber \\
\dot{n}_{+1}=\Gamma_RN_{+1}+\kappa_1p-\Gamma_{1gs}n_{+1} \nonumber
\end{eqnarray*}

where $\Delta\gamma=(\gamma_1-\gamma_0)/2$, $\bar{\gamma}=(\gamma_0+\gamma_1)/2$, $\Gamma_{2es}>\bar{\gamma}/2$, $\Delta\kappa=\kappa_1-\kappa_0$. The effect of the pump laser is described by a matrix $\mathbf{M}_{pump}$, such that immediately after the laser pulse, the state-vector is mapped to $\mathbf{x}\rightarrow \mathbf{M}_{pump}\mathbf{x}$, $\mathbf{M}_{pump}=\mathbb{1}+P\mathbf{M'}_{pump}$, where $P<1$ is the probability of being pumped into the excited state. The spin-z preserving pump matrix is as follows.

\begin{eqnarray}
M'(i,j)=-1, if ~i=j=9,10,11 \nonumber \\
M'(3,9)=M'(4,10)=M'(5,11)=+1, \nonumber \\
M'(i,j)=0, otherwise \nonumber
\end{eqnarray}

To model the dynamics following excitation, the state-vector is projected into the Eigen-basis of the matrix $\mathbf{M}$ and the eigen-states are propagated in time by multiplying by a factor $e^{-\lambda_j t}$, such that $\mathbf{x}=\sum_j \mathbf{\tilde{x}}_j(0)e^{-\lambda_j t}$. To find the signal, the state-vector is integrated over time, such that $\int_{t_g}^{\infty}dt\mathbf{x}(t)=\sum_j \frac{1}{\lambda_j}\mathbf{\tilde{x}}_j(0)e^{-\lambda_j t_g}$, and projected back into the spin-z basis.

To do the calculation, the system starts in the spin-polarized state $n=n_{+1}=1/2$ state. To prepare the initial steady-state, the spin-pump cycle is repeated about N=100 times, at a repetition period of 12ns. Then the resulting time-dependent signal is used to calculate the measured signal. In this approach, the dynamics are efficiently calculated using a set of matrix multiplications, rather than a numerical solution to the rate equations.

\subsection{Model-A: effective in-plane B-field}

In model-A, the optical pumping is considered to be spin preserving and directly maps spin-z in the ground-state to spin-z in the excited state.
In addition, the anti-crossing of the ground and excited state is treated as a constant in-plane magnetic field component.

Firstly, since the theoretical estimate of the radiative lifetime of $11~\mathrm{\mu s}$ is long compared with the measured lifetimes of the excited state, $\gamma_0=1.01~\mathrm{ns^{-1}}$, and $\gamma_1=2.03~\mathrm{ns^{-1}}$, the lifetimes are dominated by the inter-system crossing rates, and away from the anti-crossing, the read-out contrast is determined by the ratio of the ISC times, and the gate-time. 

\begin{figure}%
\includegraphics[width=1\columnwidth]{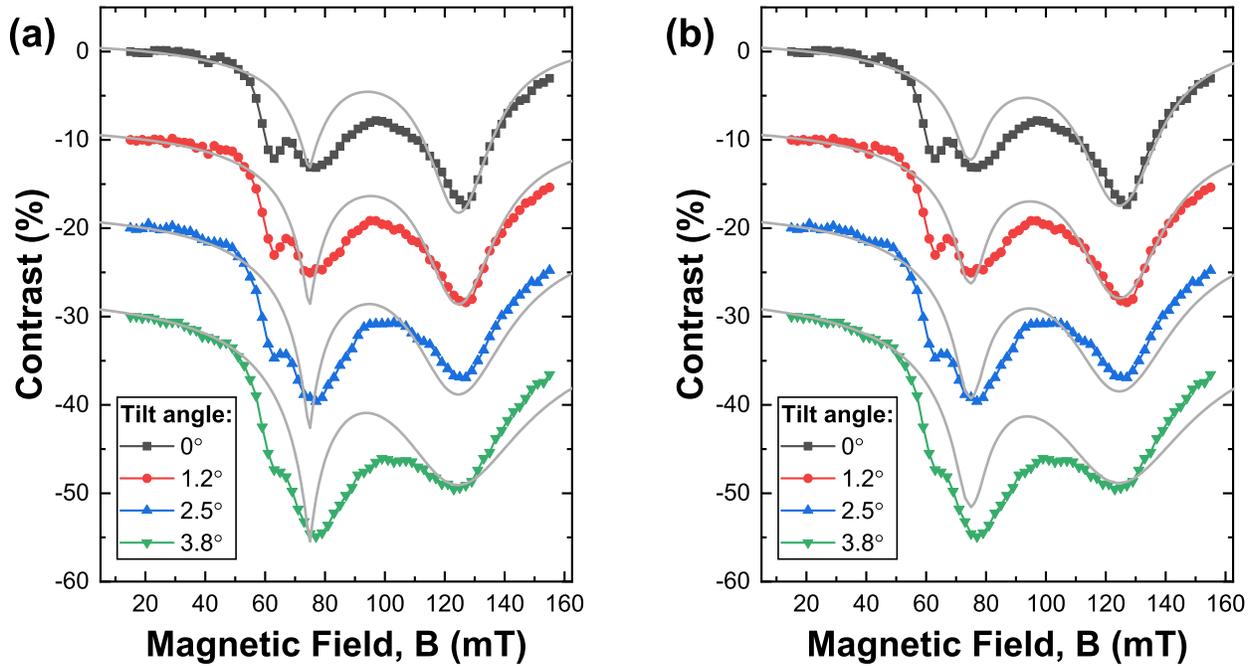}%
\caption{Plots of PL-contrast $C=\frac{PL(B)}{PL(20mT)}-1$ vs B-field, for small range of tilted B-fields. The data is reproduced from Fig. 4(c) of the main paper. In (a) the lines show calculations using Model-A. $\kappa_1/\kappa_0=0.34$, $\Omega_{y,gs}=130~\mathrm{MHz}$, $\Omega_{es,x}=140~\mathrm{MHz}$, $\kappa_0=0.02~\mathrm{ns}^{-1}.$ In (b) The lines show calculations using Model-B, where the internal magnetic fields are provided by the nuclear spins. $\kappa_1/\kappa_0=0.34$, $A=47$ MHz, and $I=3$.}
\label{fgr_SI4}%
\end{figure}

The initialization of the ground-state is largely determined by the ratio of the return inter-system crossing rates $\frac{\kappa_1}{\kappa_0}$, which is less than 1 for initialization into the $m_s=0$ state. The overall value of $\kappa_0$ is unimportant, since the contrast is a ratio, and is fixed at $\kappa_0=0.02~\mathrm{ns^{-1}}$. We have measured $T_1=10~\mathrm{\mu s}$, and a $T_2^*=19~\mathrm{ns}$ for the ground-state.  

The results of model-A are shown alongside the experimental data in Fig. \ref{fgr_SI4}(a). The GSLAC dip is relatively easy to model. It arises from the erasure of the initialized spin-z by the GSLAC. Since the read-out contrast is fixed by the ISC times $\gamma_0,\gamma_1$, the dip is determined by the spin-polarization of the initial state. This is determined by the ratio of $\kappa_1/\kappa_0\approx 0.34$, and the spin relaxation time $T_1$ which limits the build up of the spin. The strength of the effective in-plane B-field largely determines the width of the GSLAC, and fair agreement is found for a field of $\Omega_{y,gs}=130~\mathrm{MHz}$ that is aligned orthogonal to the external in-plane magnetic field direction. A co-aligned field results in a larger change in the width of the GSLAC with in-plane B-field.

The fit to the ESLAC is not so good. The ESLAC dip arises from the erasure of the read-out contrast by the ESLAC. In model-A, the laser pulse generates a spin-z in the excited state. Since $\gamma_1>\gamma_0$, the $m_s=-1$ state decays faster, resulting in a net $m_s=0$ population that is rotated out of $m_s=0$ by the effective in-plane magnetic field resulting in a darker PL. The dip is determined by the in-plane B-field and the gate-time. In order to achieve the observed change in ESLAC dip, the in-plane magnetic field needs to be co-aligned with the external in-plane B-field. This is in contrast to the ground-state. A value of $\Omega_{x,es}\approx 140~\mathrm{MHz}$ gives fair agreement of the strength of the dip. 

We note that the sensitivity to in-plane magnetic field, rules out mechanisms where the ESLAC modifies the optical pump process by degrading the spin-selectivity of the pump. Furthermore, if the sign of $D_{es}$ is negative, and the ESLAC involves the $m_s=+1$ rather than $m_s=-1$ state, this compromises the spin-selectivity and is not favored by the model. In the model, we assume the dephasing rate of the excited state is lifetime limited by the ISC-rates such that $\Gamma_{2,es}=(\gamma_0+\gamma_1)/4$. Additional pure dephasing further damps the spin-precession of the excited state, this increases the width of the ESLAC dip, but also reduces the ESLAC dip such that the model is not compatible with the observed variation with external in-plane B-field.

Model-A cannot capture the width of the ESLAC dip. However, we note that the in-plane B-field used is similar to the expected Overhauser-field for the Boron vacancy. The hyperfine constant is reported as $A=47~\mathrm{MHz}$, with the nearest neighbor nuclei being the N-14, with spin-1, yielding a total nuclear spin of $I=3$, and an Overhauser field of $141~\mathrm{MHz}$.

\subsection{Model-B: random nuclear magnetic field} 

In this model, the internal effective magnetic field is provided entirely by the hyperfine interaction, with nuclear spin I=3.
For a defect in nuclear spin-z state $m_I$, the effective B-field in z-direction is shifted to  $\Omega_z\rightarrow D-g\mu_bB_z+Am_I$. We then assume that there is a in-plane Overhauser-field of magnitude $\Omega_x=A\sqrt{I(I+1)-m_I^2}$ which is co-aligned with the external B-field for the excited state, and orthogonally aligned for the ground-state.

The ODMR spectra is calculated for each value of $-3<m_I<+3$. The ODMR is then calculated as a sum weighted by the degeneracy of the nuclear spin state. This essentially neglects nuclear spin pumping effects. As shown in Fig. \ref{fgr_SI4}(b), the width of the GSLAC is matched by the model despite no adjustable parameters that influence the GSLAC width, strongly suggesting nuclear spins provide the internal B-field. As before this is less good for larger angles.

The ESLAC peak is broadened, but retains the depth of the ESLAC peak. Given the strength of the internal field is fixed by the Hyperfine, the agreement strongly suggests the involvement of nuclear spins for the anti-crossing. The width of the ESLAC dip, in particular at B-fields between the ESLAC and GSLAC positions is not well modelled. This maybe suggests some sort of nuclear pumping effect, that is needed to rotate the Overhauser field from co-aligned (ESLAC) to cross-aligned (GSLAC) with respect to the external magnetic field.

\begin{acknowledgement}

This work was supported by the Engineering and Physical Sciences Research Council [Grant numbers EP/S001557/1, EP/026656/1, EP/L015331/1 and EP/R008809/1]. Ion implantation was performed by Keith Heasman and Julian Fletcher at the University of Surrey Ion Beam Centre. We thank W. L. Barnes for useful discussions and support.  

\end{acknowledgement}

\bibliography{hBN_ODMR}

\providecommand{\latin}[1]{#1}
\makeatletter
\providecommand{\doi}
  {\begingroup\let\do\@makeother\dospecials
  \catcode`\{=1 \catcode`\}=2 \doi@aux}
\providecommand{\doi@aux}[1]{\endgroup\texttt{#1}}
\makeatother
\providecommand*\mcitethebibliography{\thebibliography}
\csname @ifundefined\endcsname{endmcitethebibliography}
  {\let\endmcitethebibliography\endthebibliography}{}
\begin{mcitethebibliography}{35}
\providecommand*\natexlab[1]{#1}
\providecommand*\mciteSetBstSublistMode[1]{}
\providecommand*\mciteSetBstMaxWidthForm[2]{}
\providecommand*\mciteBstWouldAddEndPuncttrue
  {\def\EndOfBibitem{\unskip.}}
\providecommand*\mciteBstWouldAddEndPunctfalse
  {\let\EndOfBibitem\relax}
\providecommand*\mciteSetBstMidEndSepPunct[3]{}
\providecommand*\mciteSetBstSublistLabelBeginEnd[3]{}
\providecommand*\EndOfBibitem{}
\mciteSetBstSublistMode{f}
\mciteSetBstMaxWidthForm{subitem}{(\alph{mcitesubitemcount})}
\mciteSetBstSublistLabelBeginEnd
  {\mcitemaxwidthsubitemform\space}
  {\relax}
  {\relax}

\bibitem[Sajid \latin{et~al.}(2020)Sajid, Ford, and
  Reimers]{Sajid_IOPRepProgPhysics2020_BNreview}
Sajid,~A.; Ford,~M.~J.; Reimers,~J.~R. Single-photon emitters in hexagonal
  boron nitride: a review of progress. \textbf{2020}, \emph{83}, 044501\relax
\mciteBstWouldAddEndPuncttrue
\mciteSetBstMidEndSepPunct{\mcitedefaultmidpunct}
{\mcitedefaultendpunct}{\mcitedefaultseppunct}\relax
\EndOfBibitem
\bibitem[Tran \latin{et~al.}(2015)Tran, Bray, Ford, Toth, and
  Aharonovich]{Tran_NatNano2015}
Tran,~T.~T.; Bray,~K.; Ford,~M.~J.; Toth,~M.; Aharonovich,~I. Quantum emission
  from hexagonal boron nitride monolayers. \emph{Nature Nanotechnology}
  \textbf{2015}, \emph{11}, 37--42\relax
\mciteBstWouldAddEndPuncttrue
\mciteSetBstMidEndSepPunct{\mcitedefaultmidpunct}
{\mcitedefaultendpunct}{\mcitedefaultseppunct}\relax
\EndOfBibitem
\bibitem[Jungwirth \latin{et~al.}(2016)Jungwirth, Calderon, Ji, Spencer,
  Flatté, and Fuchs]{Jungwurth_NL2016_BNtemp}
Jungwirth,~N.~R.; Calderon,~B.; Ji,~Y.; Spencer,~M.~G.; Flatté,~M.~E.;
  Fuchs,~G.~D. Temperature Dependence of Wavelength Selectable Zero-Phonon
  Emission from Single Defects in Hexagonal Boron Nitride. \emph{Nano Letters}
  \textbf{2016}, \emph{16}, 6052--6057\relax
\mciteBstWouldAddEndPuncttrue
\mciteSetBstMidEndSepPunct{\mcitedefaultmidpunct}
{\mcitedefaultendpunct}{\mcitedefaultseppunct}\relax
\EndOfBibitem
\bibitem[Tran \latin{et~al.}(2016)Tran, Elbadawi, Totonjian, Lobo, Grosso,
  Moon, Englund, Ford, Aharonovich, and Toth]{Tran_ACSNano2016_BNRobustSPS}
Tran,~T.~T.; Elbadawi,~C.; Totonjian,~D.; Lobo,~C.~J.; Grosso,~G.; Moon,~H.;
  Englund,~D.~R.; Ford,~M.~J.; Aharonovich,~I.; Toth,~M. Robust Multicolor
  Single Photon Emission from Point Defects in Hexagonal Boron Nitride.
  \emph{ACS Nano} \textbf{2016}, \emph{10}, 7331--7338\relax
\mciteBstWouldAddEndPuncttrue
\mciteSetBstMidEndSepPunct{\mcitedefaultmidpunct}
{\mcitedefaultendpunct}{\mcitedefaultseppunct}\relax
\EndOfBibitem
\bibitem[Hoese \latin{et~al.}(2020)Hoese, Reddy, Dietrich, Koch, Fehler,
  Doherty, and Kubanek]{Hoese_SciAdv2020_FTlimited}
Hoese,~M.; Reddy,~P.; Dietrich,~A.; Koch,~M.~K.; Fehler,~K.~G.; Doherty,~M.~W.;
  Kubanek,~A. Mechanical decoupling of quantum emitters in hexagonal boron
  nitride from low-energy phonon modes. \emph{Science Advances} \textbf{2020},
  \emph{6}, eaba6038\relax
\mciteBstWouldAddEndPuncttrue
\mciteSetBstMidEndSepPunct{\mcitedefaultmidpunct}
{\mcitedefaultendpunct}{\mcitedefaultseppunct}\relax
\EndOfBibitem
\bibitem[Dietrich \latin{et~al.}(2020)Dietrich, Doherty, Aharonovich, and
  Kubanek]{Dietrich_PhysRevB2020_BN_FTlimited}
Dietrich,~A.; Doherty,~M.~W.; Aharonovich,~I.; Kubanek,~A. Solid-state single
  photon source with Fourier transform limited lines at room temperature.
  \emph{Phys. Rev. B} \textbf{2020}, \emph{101}, 081401\relax
\mciteBstWouldAddEndPuncttrue
\mciteSetBstMidEndSepPunct{\mcitedefaultmidpunct}
{\mcitedefaultendpunct}{\mcitedefaultseppunct}\relax
\EndOfBibitem
\bibitem[Awschalom \latin{et~al.}(2018)Awschalom, Hanson, Wrachtrup, and
  Zhou]{Awschalom_NatPhotonics_review}
Awschalom,~D.~D.; Hanson,~R.; Wrachtrup,~J.; Zhou,~B.~B. Quantum technologies
  with optically interfaced solid-state spins. \emph{Nature Photonics}
  \textbf{2018}, \emph{12}, 516--527\relax
\mciteBstWouldAddEndPuncttrue
\mciteSetBstMidEndSepPunct{\mcitedefaultmidpunct}
{\mcitedefaultendpunct}{\mcitedefaultseppunct}\relax
\EndOfBibitem
\bibitem[Castelletto and Boretti(2020)Castelletto, and
  Boretti]{Castelletto_IOP-JPhys_Photonics_review}
Castelletto,~S.; Boretti,~A. Silicon carbide color centers for quantum
  applications. \emph{Journal of Physics: Photonics} \textbf{2020}, \emph{2},
  022001\relax
\mciteBstWouldAddEndPuncttrue
\mciteSetBstMidEndSepPunct{\mcitedefaultmidpunct}
{\mcitedefaultendpunct}{\mcitedefaultseppunct}\relax
\EndOfBibitem
\bibitem[Gottscholl \latin{et~al.}(2020)Gottscholl, Kianinia, Soltamov,
  Orlinskii, Mamin, Bradac, Kasper, Krambrock, Sperlich, Toth, Aharonovich, and
  Dyakonov]{Gottscholl2020}
Gottscholl,~A.; Kianinia,~M.; Soltamov,~V.; Orlinskii,~S.; Mamin,~G.;
  Bradac,~C.; Kasper,~C.; Krambrock,~K.; Sperlich,~A.; Toth,~M.;
  Aharonovich,~I.; Dyakonov,~V. Initialization and read-out of intrinsic spin
  defects in a van der Waals crystal at room temperature. \emph{Nature
  Materials} \textbf{2020}, \emph{19}, 540--545\relax
\mciteBstWouldAddEndPuncttrue
\mciteSetBstMidEndSepPunct{\mcitedefaultmidpunct}
{\mcitedefaultendpunct}{\mcitedefaultseppunct}\relax
\EndOfBibitem
\bibitem[Gottscholl \latin{et~al.}(2021)Gottscholl, Diez, Soltamov, Kasper,
  Sperlich, Kianinia, Bradac, Aharonovich, and
  Dyakonov]{Gottscholl_SciAdv_VB-CoherentControl}
Gottscholl,~A.; Diez,~M.; Soltamov,~V.; Kasper,~C.; Sperlich,~A.; Kianinia,~M.;
  Bradac,~C.; Aharonovich,~I.; Dyakonov,~V. Room temperature coherent control
  of spin defects in hexagonal boron nitride. \emph{Science Advances}
  \textbf{2021}, \emph{7}, eabf3630\relax
\mciteBstWouldAddEndPuncttrue
\mciteSetBstMidEndSepPunct{\mcitedefaultmidpunct}
{\mcitedefaultendpunct}{\mcitedefaultseppunct}\relax
\EndOfBibitem
\bibitem[Gottscholl \latin{et~al.}(2021)Gottscholl, Diez, Soltamov, Kasper,
  Krau{\ss}e, Sperlich, Kianinia, Bradac, Aharonovich, and
  Dyakonov]{Gottscholl_NatComms_VB-sensing}
Gottscholl,~A.; Diez,~M.; Soltamov,~V.; Kasper,~C.; Krau{\ss}e,~D.;
  Sperlich,~A.; Kianinia,~M.; Bradac,~C.; Aharonovich,~I.; Dyakonov,~V. Spin
  defects in hBN as promising temperature, pressure and magnetic field quantum
  sensors. \emph{Nature Communications} \textbf{2021}, \emph{12}, 4480\relax
\mciteBstWouldAddEndPuncttrue
\mciteSetBstMidEndSepPunct{\mcitedefaultmidpunct}
{\mcitedefaultendpunct}{\mcitedefaultseppunct}\relax
\EndOfBibitem
\bibitem[Liu \latin{et~al.}(2021)Liu, Li, Yang, Yu, Meng, Wang, Guo, Yan, Li,
  Wang, Xu, Dong, Chen, Sun, Wang, Tang, Li, and Guo]{Liu_arXiv_VB-Rabi}
Liu,~W. \latin{et~al.}  Rabi oscillation of V$_\text{B}^-$ spin in hexagonal
  boron nitride. \emph{arXiv:2101.11220} \textbf{2021}, \relax
\mciteBstWouldAddEndPunctfalse
\mciteSetBstMidEndSepPunct{\mcitedefaultmidpunct}
{}{\mcitedefaultseppunct}\relax
\EndOfBibitem
\bibitem[Gao \latin{et~al.}(2021)Gao, Jiang, Llacsahuanga~Allcca, Shen, Sadi,
  Solanki, Ju, Xu, Upadhyaya, Chen, Bhave, and Li]{Gao_NanoLett_VB-plasmonic}
Gao,~X.; Jiang,~B.; Llacsahuanga~Allcca,~A.~E.; Shen,~K.; Sadi,~M.~A.;
  Solanki,~A.~B.; Ju,~P.; Xu,~Z.; Upadhyaya,~P.; Chen,~Y.~P.; Bhave,~S.~A.;
  Li,~T. High-Contrast Plasmonic-Enhanced Shallow Spin Defects in Hexagonal
  Boron Nitride for Quantum Sensing. \emph{Nano Letters} \textbf{2021},
  \emph{21}, 7708--7714\relax
\mciteBstWouldAddEndPuncttrue
\mciteSetBstMidEndSepPunct{\mcitedefaultmidpunct}
{\mcitedefaultendpunct}{\mcitedefaultseppunct}\relax
\EndOfBibitem
\bibitem[Kianinia \latin{et~al.}(2020)Kianinia, White, Fröch, Bradac, and
  Aharonovich]{Kianinia_ACSPhotonics_IonImplantation}
Kianinia,~M.; White,~S.; Fröch,~J.~E.; Bradac,~C.; Aharonovich,~I. Generation
  of Spin Defects in Hexagonal Boron Nitride. \emph{ACS Photonics}
  \textbf{2020}, \emph{7}, 2147--2152\relax
\mciteBstWouldAddEndPuncttrue
\mciteSetBstMidEndSepPunct{\mcitedefaultmidpunct}
{\mcitedefaultendpunct}{\mcitedefaultseppunct}\relax
\EndOfBibitem
\bibitem[Gao \latin{et~al.}(2021)Gao, Pandey, Kianinia, Ahn, Ju, Aharonovich,
  Shivaram, and Li]{Gao_ACSPhotonics_LaserWriting}
Gao,~X.; Pandey,~S.; Kianinia,~M.; Ahn,~J.; Ju,~P.; Aharonovich,~I.;
  Shivaram,~N.; Li,~T. Femtosecond Laser Writing of Spin Defects in Hexagonal
  Boron Nitride. \emph{ACS Photonics} \textbf{2021}, \emph{8}, 994--1000\relax
\mciteBstWouldAddEndPuncttrue
\mciteSetBstMidEndSepPunct{\mcitedefaultmidpunct}
{\mcitedefaultendpunct}{\mcitedefaultseppunct}\relax
\EndOfBibitem
\bibitem[Liu \latin{et~al.}(2021)Liu, Li, Yang, Yu, Meng, Wang, Li, Guo, Yan,
  Li, Wang, Xu, Wang, Tang, Li, and Guo]{Liu_ACSPhotonics2021_VBtemperature}
Liu,~W. \latin{et~al.}  Temperature-Dependent Energy-Level Shifts of Spin
  Defects in Hexagonal Boron Nitride. \emph{ACS Photonics} \textbf{2021},
  \emph{8}, 1889--1895\relax
\mciteBstWouldAddEndPuncttrue
\mciteSetBstMidEndSepPunct{\mcitedefaultmidpunct}
{\mcitedefaultendpunct}{\mcitedefaultseppunct}\relax
\EndOfBibitem
\bibitem[Chejanovsky \latin{et~al.}(2021)Chejanovsky, Mukherjee, Geng, Chen,
  Kim, Denisenko, Finkler, Taniguchi, Watanabe, Dasari, Auburger, Gali, Smet,
  and Wrachtrup]{Chejanovsky_NatMaterials_BNsingleDefect}
Chejanovsky,~N.; Mukherjee,~A.; Geng,~J.; Chen,~Y.-C.; Kim,~Y.; Denisenko,~A.;
  Finkler,~A.; Taniguchi,~T.; Watanabe,~K.; Dasari,~D. B.~R.; Auburger,~P.;
  Gali,~A.; Smet,~J.~H.; Wrachtrup,~J. Single-spin resonance in a van der Waals
  embedded paramagnetic defect. \emph{Nature Materials} \textbf{2021},
  \emph{20}, 1079--1084\relax
\mciteBstWouldAddEndPuncttrue
\mciteSetBstMidEndSepPunct{\mcitedefaultmidpunct}
{\mcitedefaultendpunct}{\mcitedefaultseppunct}\relax
\EndOfBibitem
\bibitem[Stern \latin{et~al.}(2021)Stern, Jarman, Gu, Barker, Mendelson, Chugh,
  Schott, Tan, Sirringhaus, Aharonovich, and Atatüre]{stern_arXiv_BNsingle}
Stern,~H.~L.; Jarman,~J.; Gu,~Q.; Barker,~S.~E.; Mendelson,~N.; Chugh,~D.;
  Schott,~S.; Tan,~H.~H.; Sirringhaus,~H.; Aharonovich,~I.; Atatüre,~M.
  Room-temperature optically detected magnetic resonance of single defects in
  hexagonal boron nitride. \emph{arXiv:2103.16494} \textbf{2021}, \relax
\mciteBstWouldAddEndPunctfalse
\mciteSetBstMidEndSepPunct{\mcitedefaultmidpunct}
{}{\mcitedefaultseppunct}\relax
\EndOfBibitem
\bibitem[Reimers \latin{et~al.}(2020)Reimers, Shen, Kianinia, Bradac,
  Aharonovich, Ford, and Piecuch]{Reimers_PhysRevB_VB-}
Reimers,~J.~R.; Shen,~J.; Kianinia,~M.; Bradac,~C.; Aharonovich,~I.;
  Ford,~M.~J.; Piecuch,~P. Photoluminescence, photophysics, and photochemistry
  of the ${{\mathrm{V}}_{\mathrm{B}}}^{\ensuremath{-}}$ defect in hexagonal
  boron nitride. \emph{Phys. Rev. B} \textbf{2020}, \emph{102}, 144105\relax
\mciteBstWouldAddEndPuncttrue
\mciteSetBstMidEndSepPunct{\mcitedefaultmidpunct}
{\mcitedefaultendpunct}{\mcitedefaultseppunct}\relax
\EndOfBibitem
\bibitem[Batalov \latin{et~al.}(2008)Batalov, Zierl, Gaebel, Neumann, Chan,
  Balasubramanian, Hemmer, Jelezko, and Wrachtrup]{Batalov_PRL_2008_NVTRPL}
Batalov,~A.; Zierl,~C.; Gaebel,~T.; Neumann,~P.; Chan,~I.-Y.;
  Balasubramanian,~G.; Hemmer,~P.~R.; Jelezko,~F.; Wrachtrup,~J. Temporal
  Coherence of Photons Emitted by Single Nitrogen-Vacancy Defect Centers in
  Diamond Using Optical Rabi-Oscillations. \emph{Phys. Rev. Lett.}
  \textbf{2008}, \emph{100}, 077401\relax
\mciteBstWouldAddEndPuncttrue
\mciteSetBstMidEndSepPunct{\mcitedefaultmidpunct}
{\mcitedefaultendpunct}{\mcitedefaultseppunct}\relax
\EndOfBibitem
\bibitem[Klimov \latin{et~al.}(2015)Klimov, Falk, Christle, Dobrovitski, and
  Awschalom]{Klimov_SciAdv_SiC}
Klimov,~P.~V.; Falk,~A.~L.; Christle,~D.~J.; Dobrovitski,~V.~V.;
  Awschalom,~D.~D. Quantum entanglement at ambient conditions in a macroscopic
  solid-state spin ensemble. \emph{Science Advances} \textbf{2015},
  \emph{1}\relax
\mciteBstWouldAddEndPuncttrue
\mciteSetBstMidEndSepPunct{\mcitedefaultmidpunct}
{\mcitedefaultendpunct}{\mcitedefaultseppunct}\relax
\EndOfBibitem
\bibitem[Barry \latin{et~al.}(2020)Barry, Schloss, Bauch, Turner, Hart, Pham,
  and Walsworth]{Barry_RevModPhys_NVMagnetometry}
Barry,~J.~F.; Schloss,~J.~M.; Bauch,~E.; Turner,~M.~J.; Hart,~C.~A.;
  Pham,~L.~M.; Walsworth,~R.~L. Sensitivity optimization for NV-diamond
  magnetometry. \emph{Rev. Mod. Phys.} \textbf{2020}, \emph{92}, 015004\relax
\mciteBstWouldAddEndPuncttrue
\mciteSetBstMidEndSepPunct{\mcitedefaultmidpunct}
{\mcitedefaultendpunct}{\mcitedefaultseppunct}\relax
\EndOfBibitem
\bibitem[Fuchs \latin{et~al.}(2010)Fuchs, Dobrovitski, Toyli, Heremans, Weis,
  Schenkel, and Awschalom]{Fuchs_NatPhys_NVESRabi}
Fuchs,~G.~D.; Dobrovitski,~V.~V.; Toyli,~D.~M.; Heremans,~F.~J.; Weis,~C.~D.;
  Schenkel,~T.; Awschalom,~D.~D. Excited-state spin coherence of a single
  nitrogen--vacancy centre in diamond. \emph{Nature Physics} \textbf{2010},
  \emph{6}, 668--672\relax
\mciteBstWouldAddEndPuncttrue
\mciteSetBstMidEndSepPunct{\mcitedefaultmidpunct}
{\mcitedefaultendpunct}{\mcitedefaultseppunct}\relax
\EndOfBibitem
\bibitem[Johansson \latin{et~al.}(2012)Johansson, Nation, and
  Nori]{JOHANSSON20121760}
Johansson,~J.; Nation,~P.; Nori,~F. QuTiP: An open-source Python framework for
  the dynamics of open quantum systems. \emph{Computer Physics Communications}
  \textbf{2012}, \emph{183}, 1760 -- 1772\relax
\mciteBstWouldAddEndPuncttrue
\mciteSetBstMidEndSepPunct{\mcitedefaultmidpunct}
{\mcitedefaultendpunct}{\mcitedefaultseppunct}\relax
\EndOfBibitem
\bibitem[Johansson \latin{et~al.}(2013)Johansson, Nation, and
  Nori]{JOHANSSON20131234}
Johansson,~J.; Nation,~P.; Nori,~F. QuTiP 2: A Python framework for the
  dynamics of open quantum systems. \emph{Computer Physics Communications}
  \textbf{2013}, \emph{184}, 1234 -- 1240\relax
\mciteBstWouldAddEndPuncttrue
\mciteSetBstMidEndSepPunct{\mcitedefaultmidpunct}
{\mcitedefaultendpunct}{\mcitedefaultseppunct}\relax
\EndOfBibitem
\bibitem[Fuchs \latin{et~al.}(2008)Fuchs, Dobrovitski, Hanson, Batra, Weis,
  Schenkel, and Awschalom]{Fuchs_ES_NV_PRL}
Fuchs,~G.~D.; Dobrovitski,~V.~V.; Hanson,~R.; Batra,~A.; Weis,~C.~D.;
  Schenkel,~T.; Awschalom,~D.~D. Excited-State Spectroscopy Using Single Spin
  Manipulation in Diamond. \emph{Phys. Rev. Lett.} \textbf{2008}, \emph{101},
  117601\relax
\mciteBstWouldAddEndPuncttrue
\mciteSetBstMidEndSepPunct{\mcitedefaultmidpunct}
{\mcitedefaultendpunct}{\mcitedefaultseppunct}\relax
\EndOfBibitem
\bibitem[Hanson \latin{et~al.}(2006)Hanson, Mendoza, Epstein, and
  Awschalom]{Hanson_PhysRevLett_NV-N-CoupledSpins}
Hanson,~R.; Mendoza,~F.~M.; Epstein,~R.~J.; Awschalom,~D.~D. Polarization and
  Readout of Coupled Single Spins in Diamond. \emph{Phys. Rev. Lett.}
  \textbf{2006}, \emph{97}, 087601\relax
\mciteBstWouldAddEndPuncttrue
\mciteSetBstMidEndSepPunct{\mcitedefaultmidpunct}
{\mcitedefaultendpunct}{\mcitedefaultseppunct}\relax
\EndOfBibitem
\bibitem[Anishchik and Ivanov(2017)Anishchik, and
  Ivanov]{Anishchik_PRB_NVP1_2017}
Anishchik,~S.~V.; Ivanov,~K.~L. Sensitive detection of level anticrossing
  spectra of nitrogen vacancy centers in diamond. \emph{Phys. Rev. B}
  \textbf{2017}, \emph{96}, 115142\relax
\mciteBstWouldAddEndPuncttrue
\mciteSetBstMidEndSepPunct{\mcitedefaultmidpunct}
{\mcitedefaultendpunct}{\mcitedefaultseppunct}\relax
\EndOfBibitem
\bibitem[Mendelson \latin{et~al.}(2021)Mendelson, Chugh, Reimers, Cheng,
  Gottscholl, Long, Mellor, Zettl, Dyakonov, Beton, Novikov, Jagadish, Tan,
  Ford, Toth, Bradac, and Aharonovich]{Mendelson_NatMater2021_Carbon}
Mendelson,~N. \latin{et~al.}  Identifying carbon as the source of visible
  single-photon emission from hexagonal boron nitride. \emph{Nature Materials}
  \textbf{2021}, \emph{20}, 321--328\relax
\mciteBstWouldAddEndPuncttrue
\mciteSetBstMidEndSepPunct{\mcitedefaultmidpunct}
{\mcitedefaultendpunct}{\mcitedefaultseppunct}\relax
\EndOfBibitem
\bibitem[Epstein \latin{et~al.}(2005)Epstein, Mendoza, Kato, and
  Awschalom]{Epstein_NatPhys2005_Anisotropic}
Epstein,~R.~J.; Mendoza,~F.~M.; Kato,~Y.~K.; Awschalom,~D.~D. Anisotropic
  interactions of a single spin and dark-spin spectroscopy in diamond.
  \emph{Nature Physics} \textbf{2005}, \emph{1}, 94--98\relax
\mciteBstWouldAddEndPuncttrue
\mciteSetBstMidEndSepPunct{\mcitedefaultmidpunct}
{\mcitedefaultendpunct}{\mcitedefaultseppunct}\relax
\EndOfBibitem
\bibitem[Jacques \latin{et~al.}(2009)Jacques, Neumann, Beck, Markham, Twitchen,
  Meijer, Kaiser, Balasubramanian, Jelezko, and Wrachtrup]{Jaques_PRL_NP_2009}
Jacques,~V.; Neumann,~P.; Beck,~J.; Markham,~M.; Twitchen,~D.; Meijer,~J.;
  Kaiser,~F.; Balasubramanian,~G.; Jelezko,~F.; Wrachtrup,~J. Dynamic
  Polarization of Single Nuclear Spins by Optical Pumping of Nitrogen-Vacancy
  Color Centers in Diamond at Room Temperature. \emph{Phys. Rev. Lett.}
  \textbf{2009}, \emph{102}, 057403\relax
\mciteBstWouldAddEndPuncttrue
\mciteSetBstMidEndSepPunct{\mcitedefaultmidpunct}
{\mcitedefaultendpunct}{\mcitedefaultseppunct}\relax
\EndOfBibitem
\bibitem[Fischer \latin{et~al.}(2013)Fischer, Jarmola, Kehayias, and
  Budker]{Fischer_PRB_NPatESLAC_2013}
Fischer,~R.; Jarmola,~A.; Kehayias,~P.; Budker,~D. Optical polarization of
  nuclear ensembles in diamond. \emph{Phys. Rev. B} \textbf{2013}, \emph{87},
  125207\relax
\mciteBstWouldAddEndPuncttrue
\mciteSetBstMidEndSepPunct{\mcitedefaultmidpunct}
{\mcitedefaultendpunct}{\mcitedefaultseppunct}\relax
\EndOfBibitem
\bibitem[Smeltzer \latin{et~al.}(2009)Smeltzer, McIntyre, and
  Childress]{Smeltzer_PRA_NP}
Smeltzer,~B.; McIntyre,~J.; Childress,~L. Robust control of individual nuclear
  spins in diamond. \emph{Phys. Rev. A} \textbf{2009}, \emph{80}, 050302\relax
\mciteBstWouldAddEndPuncttrue
\mciteSetBstMidEndSepPunct{\mcitedefaultmidpunct}
{\mcitedefaultendpunct}{\mcitedefaultseppunct}\relax
\EndOfBibitem
\bibitem[Falk \latin{et~al.}(2015)Falk, Klimov, Iv\'ady, Sz\'asz, Christle,
  Koehl, Gali, and Awschalom]{Falk_PRL_SiC}
Falk,~A.~L.; Klimov,~P.~V.; Iv\'ady,~V.; Sz\'asz,~K.; Christle,~D.~J.;
  Koehl,~W.~F.; Gali,~A.; Awschalom,~D.~D. Optical Polarization of Nuclear
  Spins in Silicon Carbide. \emph{Phys. Rev. Lett.} \textbf{2015}, \emph{114},
  247603\relax
\mciteBstWouldAddEndPuncttrue
\mciteSetBstMidEndSepPunct{\mcitedefaultmidpunct}
{\mcitedefaultendpunct}{\mcitedefaultseppunct}\relax
\EndOfBibitem
\end{mcitethebibliography}

\end{document}